\documentclass{aa}

\usepackage{graphicx}

\usepackage[varg]{txfonts}

\usepackage{natbib}

\begin{document} 

\title{Five transiting hot Jupiters discovered using WASP-South, Euler and
TRAPPIST: WASP-119\,b, WASP-124\,b, WASP-126\,b, WASP-129\,b and WASP-133\,b
\thanks{The photometric data and radial velocity measurements used in this
study are available at the CDS via anonymous ftp to cdsarc.u-strasbg.fr
(130.79.128.5) or via http://cdsweb.u-strasbg.fr/cgi- bin/qcat?J/A+A/.}}

\author{P. F. L. Maxted\inst{\ref{keele}} 
\and D.~R.~Anderson\inst{\ref{keele}} 
\and A.~Collier Cameron\inst{\ref{st-and}}
\and L.~Delrez\inst{\ref{liege}}
\and M.~Gillon\inst{\ref{liege}}
\and C.~Hellier\inst{\ref{keele}} 
\and E.~Jehin\inst{\ref{liege}} 
\and M.~Lendl\inst{\ref{graz},\ref{geneva}} 
\and M.~Neveu-VanMalle\inst{\ref{geneva},\ref{cambridge}}
\and F.~Pepe\inst{\ref{geneva}}  
\and D.~Pollacco\inst{\ref{warwick}}
\and D.~Queloz\inst{\ref{cambridge},\ref{geneva}} 
\and D.~S\'egransan\inst{\ref{geneva}} 
\and B.~Smalley\inst{\ref{keele}} 
\and A.~M.~S.~Smith\inst{\ref{keele},\ref{berlin}} 
\and J.~Southworth\inst{\ref{keele}} 
\and A.~H.~M.~J.~Triaud\inst{\ref{geneva},\ref{inst8},\ref{inst9}} 
\and S.~Udry\inst{\ref{geneva}}
\and T.~Wagg\inst{\ref{keele}} 
\and R.~G. West\inst{\ref{warwick}}
}
\institute{
Astrophysics Group, Keele University, Staffordshire, ST5 5BG, UK
\email{p.maxted@keele.ac.uk}
\label{keele}
\and
SUPA, School of Physics and Astronomy, University of St.\ Andrews, North
Haugh,  Fife, KY16 9SS, UK\label{st-and}
\and
Institut d'Astrophysique et de G\'eophysique, Universit\'e de
Li\`ege, All\'ee du 6 Ao\^ut, 17, Bat. B5C, Li\`ege 1, Belgium\label{liege}
\and
Space Research Institute, Austrian Academy of Sciences, Schmiedlstr. 6, 8042
Graz, Austria
\label{graz}
\and
Observatoire astronomique de l'Universit\'e de Gen\`eve
51 ch. des Maillettes, 1290 Sauverny, Switzerland\label{geneva}
\and
Cavendish Laboratory, J J Thomson Avenue, Cambridge, CB3 0HE,
UK\label{cambridge}
\and
Department of Physics, University of Warwick, Coventry CV4 7AL,
UK\label{warwick}
\and
Institute of Planetary Research, German Aerospace Center, Rutherfordstrasse 2,
12489 Berlin, Germany\label{berlin}
\and
Centre for Planetary Sciences, University of Toronto at Scarborough, 1265
Military Trail, Toronto, ON, M1C 1A4, Canada \label{inst8}
\and
Department of Astronomy \& Astrophysics, University of Toronto, Toronto, ON,
M5S 3H4, Canada\label{inst9}
}

\authorrunning{Maxted et al.}
\titlerunning{Five transiting hot Jupiters}

\abstract
{We have used photometry from the WASP-South instrument to identify 5
stars showing planet-like transits in their light curves. The planetary nature
of the companions to these stars has been confirmed using photometry from
the EulerCam instrument on the Swiss Euler 1.2-m telescope and the TRAPPIST
telescope, and spectroscopy obtained with the CORALIE spectrograph. The
planets discovered are hot Jupiter systems with orbital
periods in the range 2.17 to 5.75 days, masses from 0.3~$M_{\rm Jup}$ to
1.2~$M_{\rm Jup}$ and with radii from    1~$R_{\rm Jup}$ to  1.5~$R_{\rm
Jup}$. These planets orbit bright stars (V = 11\,--\,13) with spectral types
in the range F9 to G4. WASP-126 is the brightest planetary system in this
sample and hosts a low-mass planet with a large radius ($0.3\,M_{\rm
Jup},0.95R_{\rm Jup}$), making it a good target for transmission spectroscopy.
The high density of WASP-129\,A suggests that it is a helium-rich star similar
to HAT-P-11\,A. WASP-133\,A has an enhanced surface lithium abundance compared to
other old G-type stars, particularly other planet host stars. These planetary
systems are good targets for follow-up observations with ground-based and
space-based facilities to study their atmospheric and dynamical properties. }

\maketitle

\begin{table}
\caption{Log of observations. Data obtained after an upgrade to the CORALIE
spectrograph in November 2014 are treated separately to data obtained before
the upgrade.  In the final column we list either the number of observations
obtained with WASP-South, or the number of spectra obtained, or the filter
used to obtained the transit photometry. The different filters used are
described in the text.
\protect\rule[-1.5mm]{0mm}{2mm}}  
\begin{tabular}{lll}
\hline 
\hline 
Facility & Date &  Notes \\ [0.5mm] \hline
\multicolumn{3}{l}{{\bf WASP-119}}\\
WASP-South & 2010 Aug -- 2012 Feb& 14\,010 points \\
CORALIE  & 2013 Sep--2014 Nov &   18 spectra \\
CORALIE  & 2015 Jan--2015 Sep &    5 spectra \\
TRAPPIST & 2013 Nov 15 & NIR \\ [0.5mm]
\multicolumn{3}{l}{{\bf WASP-124}}\\ 
WASP-South & 2006 May -- 2011 Nov& 15\,165 points \\
CORALIE  & 2011 Dec -- 2014 Oct  &   28 spectra \\
CORALIE  & 2015 May -- 2015 Nov  &   13 spectra \\
TRAPPIST & 2012 Nov 17 & ExoPlanet-BB \\
TRAPPIST & 2013 May 28 & ExoPlanet-BB \\
TRAPPIST & 2013 Jun 07 & ExoPlanet-BB \\
TRAPPIST & 2013 Aug 17 & ExoPlanet-BB \\
TRAPPIST & 2013 Nov 06 & ExoPlanet-BB \\
Euler    & 2015 Jul 16 & NGTS filter \\ [0.5mm]
\multicolumn{3}{l}{{\bf WASP-126}}\\ 
WASP-South & 2010 Aug--2012 Feb & 13\,245 points \\
CORALIE  & 2013 Oct--2014 Oct  &   22 spectra \\
CORALIE  & 2014 Nov--2015 Sep  &    8 spectra \\
TRAPPIST & 2014 Jan 08 & $z^{\prime}$ \\
TRAPPIST & 2014 Jan 31 & $z^{\prime}$ \\
Euler & 2014 Sep 05 & Gunn $r$ filter \\
Euler & 2014 Oct 17 & Gunn $r$ filter \\
Euler & 2014 Nov 06 & Gunn $r$ filter \\ [0.5mm]
\multicolumn{3}{l}{{\bf WASP-129}}\\ 
WASP-South & 2006 May--2012 May & 26\,940 points \\
CORALIE  & 2014 Feb--2014 Jun  &   9 spectra \\
Euler & 2015 Mar 08 & NGTS filter\\
Euler & 2015 Mar 31 & I$_{\rm C}$ \\
TRAPPIST & 2015 Mar 31 & NIR \\ [0.5mm]
\multicolumn{3}{l}{{\bf WASP-133}}\\ 
WASP-South & 2006 May--2012 Jun & 18\,436 points \\
CORALIE  & 2014 Aug--2014 Oct  &   12 spectra \\
TRAPPIST & 2014 Sep 26 & ExoPlanet-BB \\
Euler & 2015 Jun 14 & NGTS filter\\
TRAPPIST & 2015 Aug 25 & ExoPlanet-BB \\
\hline 
\end{tabular} 
\end{table} 

\begin{table*}
\caption{Light curves of five transiting exoplanet systems. This table is
available in its entirety at CDS. The first few lines are shown here for
guidance as to its contents and format. Column 1 is the WASP identification
number of the star. Column 2 indicates the source of the observation as
follows: W -- WASP-South; T -- TRAPPIST; E -- EulerCam.  Column 3 indicates
the filter used for the observations as follows: WASP -- 400\,--\,700\,nm; NIR
-- Astrodon NIR 700~nm cut-on luminance; Exo -- Astrodon ExoPlanet-BB; NGTS --
EulerCam NGTS; z\_p -- $z^{\prime}$; Gunn -- Gunn R; I\_C -- Cousins I$_{\rm
C}$. Column 4 is the UTC date of the observation given as HJD-2450000. Columns
5 and 6 are the apparent magnitude of the star and its standard error relative
to the mean out-of-transit apparent magnitude. Column 7 is a flag that is set
to the value 1 for the first observation after a meridian flip with the
TRAPPIST telescope.\label{lctable}}
\begin{tabular}{lllrrrr}
\hline
\hline
\multicolumn{1}{l}{WASP-} &
\multicolumn{1}{l}{Source} &
\multicolumn{1}{l}{Filter} &
\multicolumn{1}{l}{Date} &
\multicolumn{1}{l}{$\Delta m$} &
\multicolumn{1}{l}{$\sigma_m$} &
\multicolumn{1}{l}{Flip} \\
\hline
119 & W & WASP & 5780.52513 &$ 0.0123$&  0.0163 & 0 \\
119 & W & WASP & 5780.52557 &$-0.0135$&  0.0159 & 0 \\
119 & W & WASP & 5780.54147 &$-0.0014$&  0.0156 & 0 \\
\hline
\end{tabular}
\end{table*}

\begin{table}
\caption{CORALIE radial velocities.  This table is available in its entirety at
CDS. The first few lines are shown here for guidance as to its contents and
format. Data are provided to the full precision used for calculations, but
times of mid-exposure are only accurate to a few minutes. The standard error on
the bisector span (BS) measurement is twice the standard error on the RV
($\sigma_{\rm RV}$). \label{rvtable}}
\begin{tabular}{rrrrr} 
\multicolumn{4}{l}{{\bf WASP-119}}\\  
\hline 
\multicolumn{1}{l}{WASP-} &\multicolumn{1}{l}{BJD(UTC)}
& \multicolumn{1}{l}{RV} &
\multicolumn{1}{l}{$\sigma_{\rm RV}$} & \multicolumn{1}{l}{BS} \\
& --\,2\,400\,000  & (km s$^{-1}$) & (km s$^{-1}$) & (km s$^{-1}$)\\ [0.5mm] \hline
119 & 56548.906910 & 8.45864 & 0.01532 &$-0.09451$ \\
119 & 56570.857181 & 8.24473 & 0.01756 &$ 0.01842$ \\
119 & 56575.777466 & 8.24146 & 0.01258 &$ 0.01012$ \\
119 & 56593.863562 & 8.42141 & 0.02158 &$-0.00951$ \\
119 & 56622.794968 & 8.33225 & 0.02814 &$-0.05036$ \\
\hline
\end{tabular} 
\end{table}

\begin{table}
\caption{Stellar parameters. The V magnitude is based on the estimated
apparent Gaia G magnitude from \citet{2013yCat.1324....0S} and the
transformation V$-$G from  \citet{2010A&A...523A..48J} assuming V$-$I$_{\rm
C}\approx 0.7$ for all stars. Other symbols are defined in the text. The
1SWASP identifier provides the J2000.0 coordinates of the star taken from
 the USNO-B1.0 catalogue \citep{2003AJ....125..984M}.
\label{star_table}}
\begin{tabular}{lr}
\hline
Parameter & \multicolumn{1}{l}{Value} \\
\hline
\hline
\multicolumn{2}{l}{{\bf WASP-119}}\\
Identifiers & \multicolumn{1}{l}{1SWASP\,J034343.96$-$651137.8}\\
            & \multicolumn{1}{l}{2MASS\,03434402$-$6511378} \\
Spectral type & \multicolumn{1}{l}{G5} \\
V & 12.2 $\pm$ 0.4 \\
$T_{\rm eff}$ (K)  & 5650  $\pm$ 100  \\
$\log g$ (cgs)& 4.5 $\pm$ 0.3    \\
$v\,\sin I$ (km\,s$^{-1}$)     &    0.7 $\pm$ 0.9      \\
{[Fe/H]}   &   +0.14 $\pm$ 0.10     \\
log A(Li)  &  1.36 $\pm$ 0.10  \\
\noalign{\smallskip}
\multicolumn{2}{l}{{\bf WASP-124}}\\
Identifiers & \multicolumn{1}{l}{1SWASP\,J221051.43$-$304458.3} \\
            & \multicolumn{1}{l}{2MASS\,22105143$-$3044588} \\
Spectral type & \multicolumn{1}{l}{F9} \\
V & 12.7 $\pm$ 0.4 \\
$T_{\rm eff}$ (K)  & 6050  $\pm$ 100  \\
$\log g$ (cgs)& 4.0 $\pm$ 0.25     \\
$v\,\sin I$ (km\,s$^{-1}$)     &    3.2 $\pm$ 0.9      \\
{[Fe/H]}   &  $-0.02  \pm$ 0.11     \\
log A(Li)  &  2.67 $\pm$ 0.08  \\
\noalign{\smallskip}
\multicolumn{2}{l}{{\bf WASP-126}}\\
Identifiers & \multicolumn{1}{l}{1SWASP\,J041329.75$-$691336.6}\\
            & \multicolumn{1}{l}{TYC 9153-833-1} \\
            & \multicolumn{1}{l}{2MASS\,04132972$-$6913365}\\
Spectral type & \multicolumn{1}{l}{G2} \\
V & 10.8 $\pm$ 0.5 \\
$T_{\rm eff}$ (K)  & 5800  $\pm$ 100  \\
$\log g$ (cgs)& 4.4 $\pm$ 0.25     \\
$v\,\sin I$ (km\,s$^{-1}$)     &    0.5 $\pm$ 0.5      \\
{[Fe/H]}   &   +0.17 $\pm$ 0.08     \\
log A(Li)  &  1.18 $\pm$ 0.09  \\
\noalign{\smallskip}
\multicolumn{2}{l}{{\bf WASP-129}}\\
Identifiers & \multicolumn{1}{l}{1SWASP\,J114511.75$-$420350.3} \\
            & \multicolumn{1}{l}{TYC 7749-508-1} \\
            & \multicolumn{1}{l}{2MASS\,11451175$-$4203501} \\
Spectral type & \multicolumn{1}{l}{G1} \\
V & 12.3 $\pm$ 0.6 \\
$T_{\rm eff}$ (K)  & 5900  $\pm$ 100  \\
$\log g$ (cgs)& 4.1 $\pm$ 0.25     \\
$v\,\sin I$ (km\,s$^{-1}$)     &    2.7 $\pm$ 0.6      \\
{[Fe/H]}   &   +0.15 $\pm$ 0.09     \\
log A(Li)  &  2.41 $\pm$ 0.09  \\
\noalign{\smallskip}
\multicolumn{2}{l}{{\bf WASP-133}}\\
Identifiers & \multicolumn{1}{l}{1SWASP\,J205818.07$-$354747.6} \\
            & \multicolumn{1}{l}{2MASS\,20581808$-$3547475} \\
Spectral type & \multicolumn{1}{l}{G4} \\
V & 12.9 $\pm$ 0.4 \\
$T_{\rm eff}$ (K)  & 5700  $\pm$ 100  \\
$\log g$      & 4.1 $\pm$ 0.25     \\
$v\,\sin I$ (km\,s$^{-1}$)     &    0.3 $\pm$ 0.7      \\
{[Fe/H]}   &   +0.29 $\pm$ 0.12     \\
log A(Li)  &  2.67 $\pm$ 0.09  \\
\hline
\end{tabular}
\end{table} 

\section{Introduction}

 Ground-based transit surveys such as WASP  \citep{2008MNRAS.385.1576P}, HAT
\citep{2004PASP..116..266B}, HAT-South \citep{2013PASP..125..154B} and KELT
\citep{2008AJ....135..907P} have identified the majority of bright hot Jupiter
planetary systems that have been studied in detail using observations with
large ground-based telescopes and space-based instrumentation. Although
space-based surveys such as Kepler-K2 \citep{2014PASP..126..398H}, TESS
\citep{2015JATIS...1a4003R} and PLATO \citep{2014ExA....38..249R} will produce
much higher quality photometry than is possible from the ground, it is
advantageous to identify as many planetary systems as possible in the areas of
the sky that will be observed by these missions so that the observing strategy
can be optimised. This was clearly demonstrated by the recent discovery that
WASP-47  is a very rare example of a multiple-planet system containing a hot
Jupiter \citep{2015ApJ...812L..18B, 2015arXiv150907750N}. This discovery was
only possible because this known planetary system was prioritised for
high-cadence observations with Kepler-K2. These high-cadence observations made
it possible to detect the transit time variations produced by the
gravitational interactions between the planets in this system. The discovery
that WASP-47 is a multi-planet system also demonstrates that detailed
obervations of apparently normal hot Jupiter systems continue to reveal
unexpected properties of this diverse group of planetary systems.

 Here we present the discovery of 5 hot-Jupiter planetary systems in the
southern hemisphere. Section 2 describes the observations we have obtained,
section 3 describes our analysis of the host stars, the stellar and planetary
masses and radii are derived in section 4 and we discuss the properties of
each system briefly in section 5.

\section{Observations}

 WASP-South uses an array of 8 cameras that observe selected regions of the
southern sky with a combined area of approximately 450 square degrees at a
typical cadence of about 8 minutes and the exposure time is 30\,s. 
The WASP survey and instruments are
described in \citet{2006PASP..118.1407P} and a description of our
planet-hunting methods can be found in \citet{2008MNRAS.385.1576P} and
\citet{2007MNRAS.375..951C}.

 WASP-South planet candidates are followed up using the TRAPPIST robotic
photometer \citep{2011Msngr.145....2J,2011EPJWC..1106002G}, and the CORALIE
spectrograph and EulerCam photometer \citep{2012A&A...544A..72L} on the Swiss
Euler 1.2-m telescope at La~Silla.  We find that about 1-in-12 candidates turns
out to be a planet, the remainder being blends that are unresolved in the WASP
images (which have 14 arcsec pixels) or astrophysical transit mimics, usually
eclipsing binary stars.   A list of observations reported in this paper is
given in Table~1. The light curve data used in our analysis are given in
Table~\ref{lctable}  and the radial velocities  measured from the CORALIE
spectra are listed in Table~\ref{rvtable}.  The radial velocity measurements
presented in Table~\ref{rvtable} use a mask based on a G2-type template for
the calculation of the cross-correlation function. We also measured the radial
velocity using a K5-type  template and found that the resuls are consistent to
within 7\,m\,s$^{-1}$. This test excludes most scenarios in which the periodic
RV variations we observe are due to an eclipsing binary star blended with the
light of the target star \citep{2008A&A...489L...9H}. Also given in
Table~\ref{rvtable} are the bisector span (BS) measurements that characterise
the asymmetry in the stellar line profiles for each spectrum
\citep{2001A+A...379..279Q}. The bisector span will show a significant
correlation with the measured radial velocity (RV) if the apparent variations
in RV are due to stellar magnetic activity (star spots) or blended spectra in
a multiple star system. We used the Bayesian information criterion (BIC) to
test the significance of any correlation between RV and BS for each star,
i.e., we calculated the change of BIC going from a constant BS value (weighted
mean)  to a fit of a straight line to BS as a function of RV. The zero-points
of the RV and BS measurements may be slightly different for spectra obtained
before or after the upgrade to CORALIE in November 2014 so we analysed data
either side of this date independently. Weak evidence for a correlation is
present in the data obtained prior to the CORALIE upgrade for WASP-124 (BIC
decreased by 6.6) but this apparent correlation is not confirmed by the data
obtained post-upgrade and there is no other indication that WASP-124 is an
active star or a multiple star system, so we assume that this correlation is a
statistical fluke. None of the data sets for any other stars show any evidence
for a correlation between BS and RV.

To measure precise planetary radii, we obtained follow-up  photometry of the
transits of these stars with EulerCam and TRAPPIST. The dates of observation
and filters used are given in Table 1. The NGTS filter on the EulerCam
instrument has a central wavelength of 698\,nm and the effective bandwidth is
312\,nm. The ExoPlanet-BB filter manufactured by Astrodon blocks wavelengths
below about 500nm. The NIR filter is an NIR luminance filter
also from Astrodon that blocks light below 700~nm.

\section{The host stars} 
\subsection{Spectroscopic analysis}
 The CORALIE spectra of the host stars were co-added to produce spectra for
analysis using the methods described in \citet{2009A&A...496..259G} and
\citet{2013MNRAS.428.3164D}. We used the H$\alpha$ line to estimate the
effective temperature ($T_{\rm eff}$), and the Na~{\sc i} D and Mg~{\sc i} b
lines as diagnostics of the surface gravity ($\log g$). The resulting
parameters are listed in Table~\ref{star_table}.  The CORALIE spectra do
not have a sufficient number of clean spectral lines to determine reliable
values for [Mg/Fe] or [Na/Fe] independently of $\log g$  so the quoted error
in $\log g$ includes an additional 0.15 dex error due to the unknown Mg
abundance of these stars \citep{2015ApJ...805..126B}. The iron abundances were
determined from equivalent-width measurements of several clean and unblended
Fe~{\sc i} lines and are given relative to the solar value presented in
\citet{2009ARA&A..47..481A}. The quoted abundance errors include that given by
the uncertainties in $T_{\rm eff}$ and $\log g$, as well as the scatter due to
measurement and atomic data uncertainties. The projected rotation velocities
($v \sin I$) were determined by fitting the profiles of the Fe~{\sc i} lines
after convolving with the CORALIE instrumental resolution ($R$ = 55\,000) and
a macroturbulent velocity adopted from the calibration of
\citet{2014MNRAS.444.3592D}. 

\subsection{Rotational modulation}
 We searched the WASP photometry of each star for rotational modulations by
using a sine-wave fitting algorithm as described by
\citet{2011PASP..123..547M}. We estimated the significance of periodicities by
subtracting the fitted transit light curve and then repeatedly and randomly
permuting the nights of observation.  For none of our stars was a significant
periodicity obtained, with 95\%-confidence upper limits on the semi-amplitude
being typically 1 mmag.

\section{System parameters}
 The CORALIE radial-velocity measurements were combined with the WASP, EulerCam
and TRAPPIST photometry in a simultaneous Markov-chain Monte-Carlo (MCMC)
analysis to find the system parameters. For details of our methods see
\citet{2007MNRAS.375..951C}. Limb-darkening was parameterised using the
4-parameter non-linear law and coefficients from \citet{2000A&A...363.1081C}.
The value of T$_{\rm eff}$ used to interpolate the limb darkening coefficients
from these tables is a free parameter in the fit but is constrained by a
Gaussian prior according to value of T$_{\rm eff}$ and its standard error for
each star from Table~\ref{star_table}. The resulting values of T$_{\rm eff}$
are all found to be consistent with prior values so we do not quote them here.
We used the coefficients for the R-band to model the WASP light curves and
those obtained with the ExoPlanet-BB, NGTS and Gunn R filters. For the
z$^{\prime}$ and NIR filters we used the coefficients for the Sloan
z$^{\prime}$-band from \citet{2004A&A...428.1001C}. The coefficients for
log~g=4.5, [M/H]=0.1 and micro-turbulence velocity V$_{\rm T}=2$\,km\,s$^{-1}$
were interpolated from these tables for each trial value of the effective
temperature in the MCMC chain. Changing the choice of table used to calculate
the limb darkening coefficients assigned to these non-standard filters from
z$^{\prime}$ to  R-band or from R-band to I-band changes the mass and radius
derived for the stars and planets by less than the quoted uncertainty. For
TRAPPIST observations where a meridian flip was required the data obtained
with the telescope on the west side and east side of the German equatorial
mount were analysed independently to allow for any change in the
photometric zero-point. To account for stellar noise in the radial velocity
measurements (``jitter'') we used a circular orbit fit by least-squares with
the orbital period and time of mid-transit fixed to values determined from the
photometry. We then used trial-and-error to find the amount of jitter required
to achieve a reduced $\chi^2$ value of 1 when this additional variance is
added in quadrature to the standard errors given in Table~\ref{rvtable}. The
values of the jitter ($\sigma_{\rm jit}$) used in our analysis are given in
Table~\ref{tab:mcmc}. We did not find strong evidence  in any of these systems
for a linear drift in the apparent centre-of-mass velocity,
$\frac{d\gamma}{dt}$. There is weak evidence for a drift in the velocity of
WASP-126\,A
($\frac{d\gamma}{dt}=(5.5\pm3.0)\times10^{-5}$\,km\,s$^{-1}$\,d$^{-1}$), but
we do not consider this to be a significant detection. Adding this extra
parameter in the modelling of the data changes the other parameters by less
than one standard deviation.

 For all of our planets the data are compatible with zero eccentricity and
hence we imposed a circular orbit -- the rationale for this assumption is
discussed in \citet{2012MNRAS.422.1988A}. The fitted parameters were thus
$T_{\rm c}$, $P$, $\Delta F$, $T_{14}$, $b$, $K_{\rm 1}$, where $T_{\rm c}$ is
the epoch of mid-transit, $P$ is the orbital period, $\Delta F$ is the
planet-star area ratio (i.e.,  the fractional flux-deficit that would be
observed during transit in the absence of limb-darkening), $T_{14}$ is the
total transit duration (from first to fourth contact), $b$ is the impact
parameter of the planet's path across the stellar disc, and $K_{\rm 1}$ is the
stellar reflex velocity semi-amplitude.

\begin{figure}
\resizebox{\hsize}{!}{\includegraphics{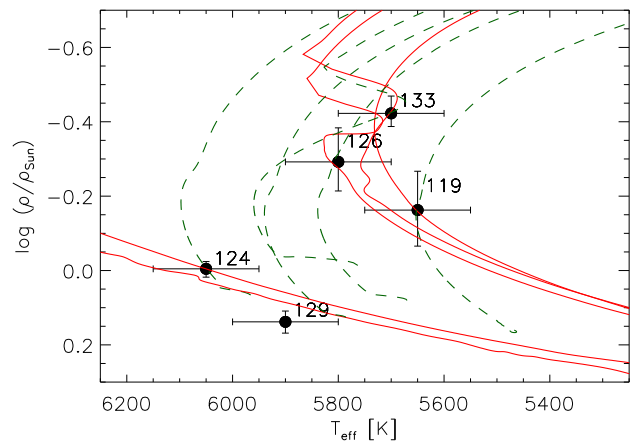}}
\caption{Planet host stars in the T$_{\rm eff}$ -- density plane (error bars
labelled by WASP identification number) compared to 
best-fit evolution tracks (dashed lines) and isochrones (solid lines) 
for the masses and ages listed in Table~\ref{bagemass_table} assuming [Fe/H] as
listed in Table~\ref{star_table}.  
\label{trho_plot}
}
\end{figure}

 The analysis of the transit light curves leads directly to an estimate of the
stellar density \citep{2003ApJ...585.1038S} and  the spectroscopic orbit can
then be used to infer the planetary surface gravity
\citep{2007MNRAS.379L..11S}. One additional constraint is required to obtain
the masses and radii of the star and planet. The additional constraint we have
used is the mass of the star estimated using stellar models based on the
effective temperature, metallicity and mean density of the star. We used the
open source software {\sc
bagemass}\footnote{\url{http://sourceforge.net/projects/bagemass}} to calculate
the posterior mass distribution for each star  using the Bayesian method
described by \citet{2015A&A...575A..36M}. The models used in {\sc bagemass}
were calculated using the {\sc garstec} stellar evolution code
\citep{2008Ap&SS.316...99W}. The mean and standard deviation of the posterior
mass distribution for the star are included in the MCMC analysis as a Gaussian
prior. The mass and age of the stars derived are shown in
Table~\ref{bagemass_table}. There is no good match to the observed density of
WASP-129\,A  if we use the standard grid of stellar models in {\sc bagemass}
with the following linear relation between helium abudance, ${\rm Y}$ and
heavy element abundance, ${\rm Z}$: ${\rm Y} = 0.2485 + 0.984{\rm Z}, $ where
the zero-point in this relation is the primordial helium abundance
\citep{2010JCAP...04..029S} and the gradient is set to match the current solar
helium abundance. For this star we used the grid of stellar models
provided within bagemass with an enhanced initial helium abundance ($\Delta
\rm Y = +0.02$). This has the effect of reducing the mass estimate for this
star by about one standard deviation. The best-fit stellar evolution tracks
and isochrones are shown in Fig.~\ref{trho_plot}. 

 The parameters for each planetary system derived from our analysis are given
in Table~\ref{tab:mcmc}. The discovery data and model fits to these
observations are shown in Figs. 2\,--\,6. 

\begin{table}
\caption{Stellar mass and age estimates. The mean and standard deviation of
the posterior distributions are given together with the best-fit values in
parentheses. \label{bagemass_table} }
\begin{tabular}{lrr}
\hline
Star & \multicolumn{1}{l}{Mass [$M_{\rm \odot}$]}& 
\multicolumn{1}{l}{Age [Gyr]}\\
\hline
\hline
WASP-119\,A & 1.02 $\pm$ 0.06 (1.01)& 8.0 $\pm$ 2.5 (8.1) \\
WASP-124\,A & 1.07 $\pm$ 0.05 (1.10)& 2.1 $\pm$ 1.4 (1.1) \\
WASP-126\,A & 1.12 $\pm$ 0.06 (1.10)& 6.4 $\pm$ 1.6 (6.4) \\
WASP-129\,A \tablefootmark{a}
         & 1.00 $\pm$ 0.03 (1.03)& 1.0 $\pm$ 0.9 (0.0) \\
WASP-133\,A & 1.16 $\pm$ 0.08 (1.20)& 6.8 $\pm$ 1.8 (5.7) \\
\hline
\end{tabular}
\tablefoot{
\tablefoottext{a}{Values for helium-enhanced models, $\Delta \rm Y = +0.02$}
}
\end{table}

\begin{figure}
\resizebox{0.95\hsize}{!}{\includegraphics{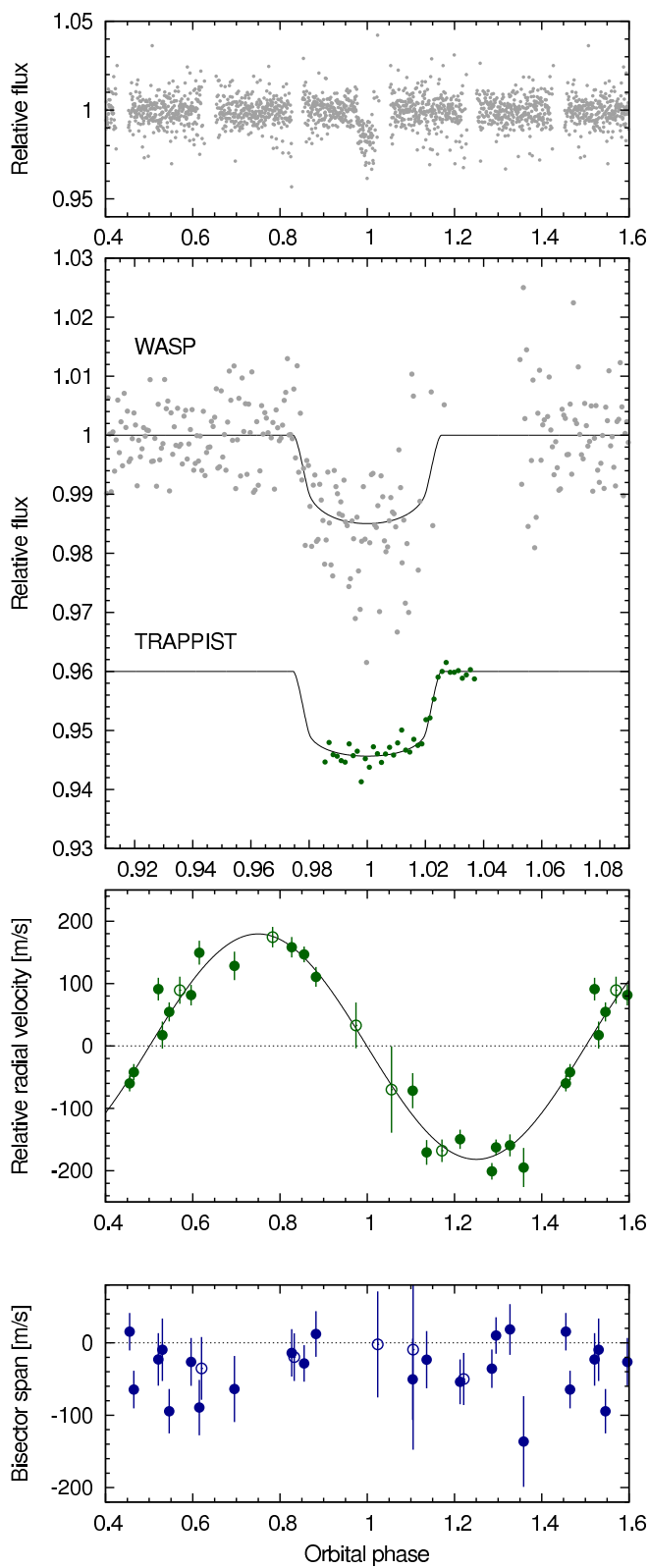}}
\caption{WASP-119\,b discovery data: (Top) The WASP-South light curve folded on
the transit period. (Second panel) The binned WASP data with (offset) the
follow-up transit light curves (ordered from top-to-bottom as in Table~1)
together with the fitted MCMC model.  (Third) The CORALIE radial velocities
with the fitted model. Filled/open symbols denote data obtained before/after
the upgrade to the CORALIE spectrograph, respectively. (Lowest) The bisector
spans; the absence of any significant correlation with radial velocity is a
check against transit mimics.}
\end{figure}

\begin{figure}
\resizebox{0.95\hsize}{!}{\includegraphics{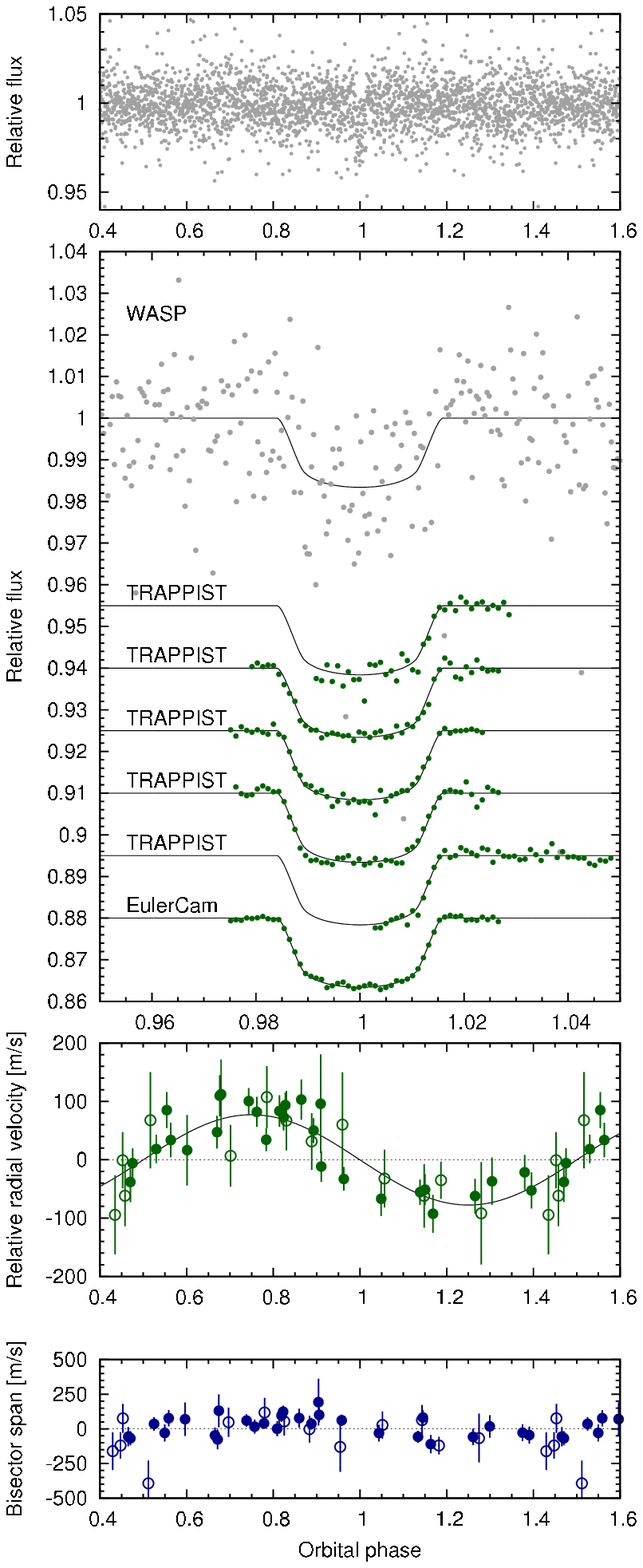}}
\caption{WASP-124\,b discovery data, as for Fig 1.}
\end{figure}

\begin{figure}
\resizebox{0.95\hsize}{!}{\includegraphics{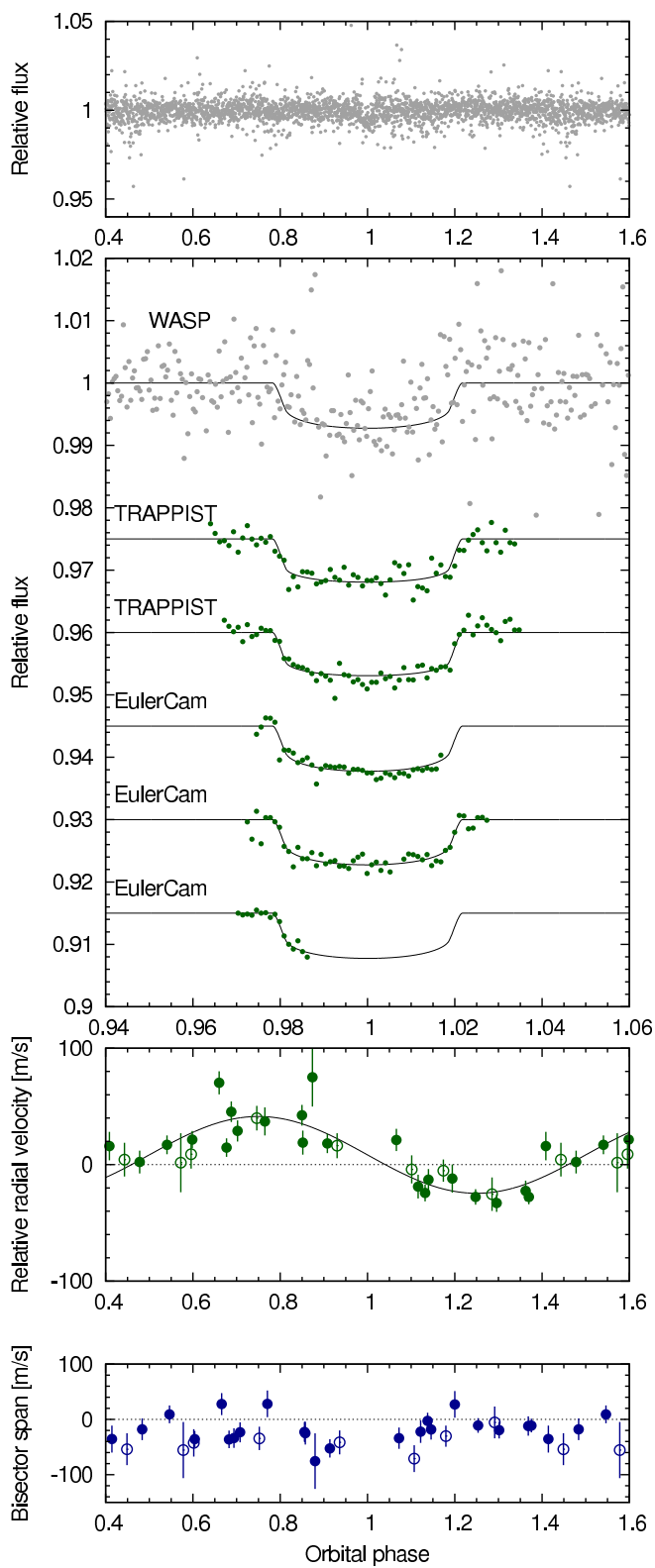}}
\caption{WASP-126\,b discovery data, as for Fig 1. The measured drift in the centre-of-mass
velocity given in Table~\ref{tab:mcmc} has  been subtracted from the
radial velocities prior to plotting.}
\end{figure}

\begin{figure}
\resizebox{0.95\hsize}{!}{\includegraphics{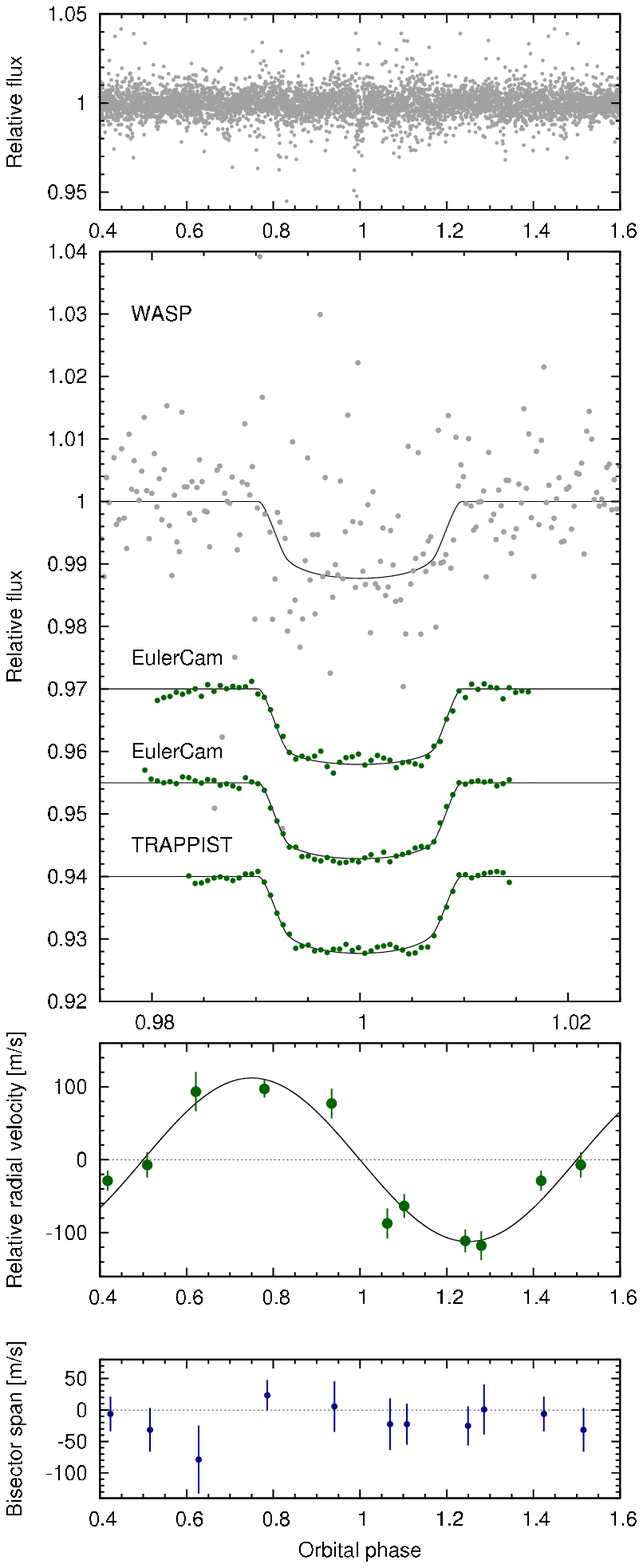}}
\caption{WASP-129\,b discovery data, as for Fig 1.}
\end{figure}

\begin{figure}
\resizebox{0.95\hsize}{!}{\includegraphics{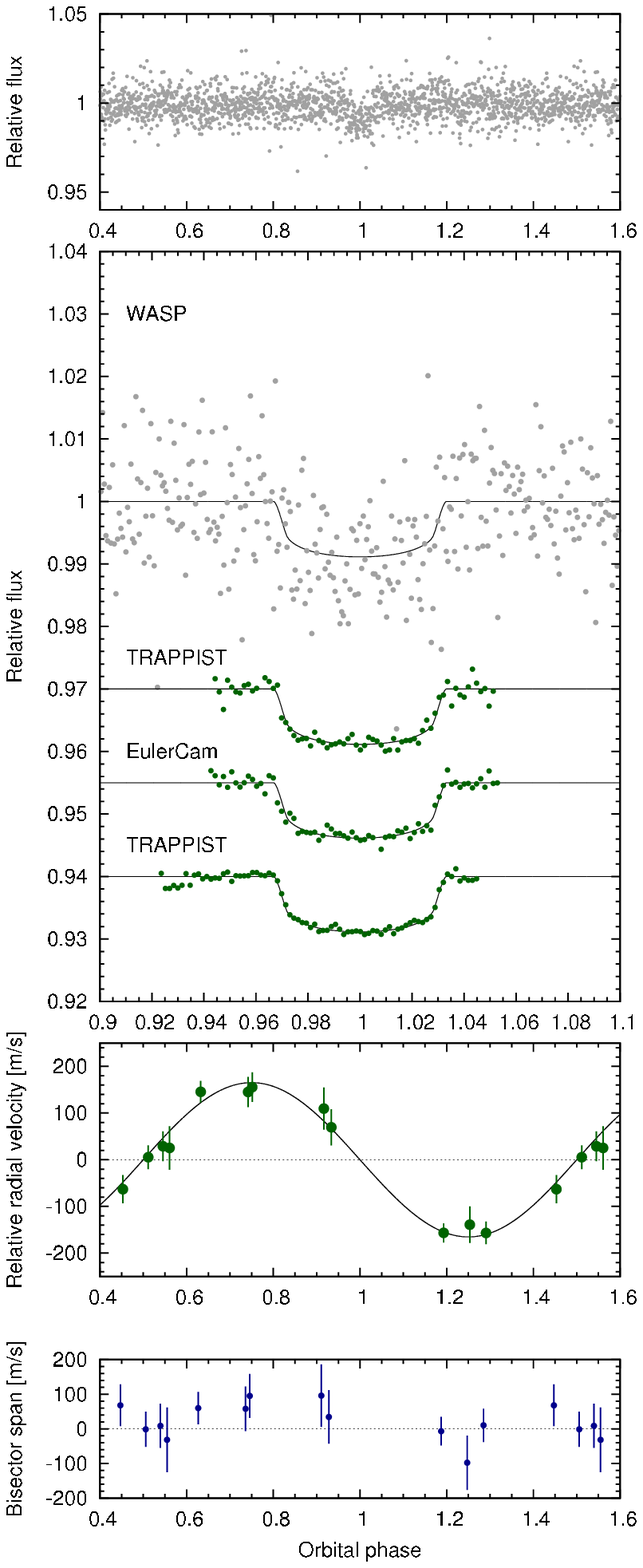}}
\caption{WASP-133\,b discovery data, as for Fig 1.}
\end{figure}

\subsection{Kinematics}
 The kinematics of our sample in Galactic coordinates are summarized in
Table~\ref{uvwtable}. The proper motions are taken from
\citet{2013yCat.1324....0S} and the radial velocities are from
Table~\ref{tab:mcmc}. To calculate the distance we use the apparent K$_{\rm
s}$-band magnitude from \citet{2006AJ....131.1163S}, the radius of the star
from Table~\ref{tab:mcmc} and the angular diameter of the star based on the
calibration of K-band surface brightness -- effective temperature relation
from \citet{2004A&A...426..297K}. We assume that interstellar reddening is
negligible and that K$ = {\rm K}_{\rm s} +0.044$ \citep{2005ARA&A..43..293B}.
The transformation of these variables to Galactic (U, V, W) velocity in the
local standard of rest is done using the procedure {\tt
gal\_uvw}.\footnote{idlastro.gsfc.nasa.gov}  U is positive towards the Galactic
anti-centre and the solar motion is taken from \citet{2011MNRAS.412.1237C}.

 \citet{2007A&A...475..519H}  find that individual stars in the solar
neighbourhood cannot be unambiguously assigned to different populations based
only on their kinematics, but the well-defined age -- velocity dispersion
relation for solar-type stars does suggest that WASP-126\,A is likely to be older
than about 4~Gyr based on its kinematics, consistent with the age of about
6~Gyr we have derived using stellar models.

\begin{table*} 
\caption{System parameters. The UTC time of mid-transit, $T_{\rm c}$, is given
as HJD$-$2450000. Upper limits on the eccentricity, $e$, are 95\% confidence
limits. Values in parantheses are standard errors in the final digit.$T_{\rm
eq}$ is the planet equilibrium temperature assuming zero albedo. Where two
values of the jitter ($\sigma_{\rm jit}$) are given, these apply to radial
velocity measurements taken before and after the upgrade to the CORALIE
spectrograph, respectively.} 
\label{tab:mcmc}
\begin{tabular}{lrrrrr}
\hline
Parameter                           & WASP-119              & WASP-124                & WASP-126                & WASP-129                & WASP-133 \\ 
\hline 
\hline 
$P$ (d)                             & 2.49979(1)            & 3.372650(1)             & 3.28880(1)              & 5.748145(4)             & 2.176423(1)             \\
$T_{\rm c}$                         & 6537.547(2)           & 7028.5834(1)            &  6890.3191(6)           & 7027.4373(2)            & 7116.9874(2)            \\
$T_{\rm 14}$ (d)                    & 0.126 $\pm$ 0.006     & 0.1071 $\pm$ 0.0006     &  0.142 $\pm$ 0.002      & 0.1119 $\pm$ 0.0008     & 0.1433 $\pm$ 0.0007     \\
$\Delta F$                          & 0.0131 $\pm$ 0.0008   & 0.0154 $\pm$ 0.0002     & 0.0061 $\pm$ 0.0002     & 0.0114 $\pm$ 0.0001     & 0.0074 $\pm$ 0.0001     \\
$b$                                 & 0.50$^{+0.10}_{-0.24}$& 0.61 $\pm$ 0.02         & 0.3 $\pm$ 0.2           & 0.62 $\pm$ 0.02         & 0.2  $\pm$ 0.1          \\ \noalign{\smallskip}
$i$ ($^\circ$)                      & 85 $\pm$ 2            & 86.3 $\pm$ 0.2          & 87.9$\pm$ 1.5           & 87.7 $\pm$ 0.2          & 87 $\pm$ 1              \\
$R_{\star}/a$                       & $0.158 \pm 0.015$     & $0.106 \pm 0.018$      & $0.131^{+0.011}_{-0.004}$& $0.066 \pm 0.0017$   & $0.194 ^{+0.006}_{-0.003}$ \\
$K_{\rm 1}$ (km s$^{-1}$)           & 0.181 $\pm$ 0.010     & 0.077 $\pm$ 0.008       & 0.036 $\pm$ 0.005       & 0.11 $\pm$ 0.01         & 0.17 $\pm$ 0.01         \\
$\gamma$ (km s$^{-1}$)              & 8.404 $\pm$ 0.007     &$-5.876 \pm$ 0.006       & 29.164 $\pm$ 0.004      & 21.988 $\pm$ 0.008      & $-23.913 \pm$ 0.009     \\
$M_{\star}$ ($M_{\rm \odot}$)       & 1.02 $\pm$ 0.06       & 1.07 $\pm$ 0.05         & 1.12 $\pm$ 0.06         & 1.00 $\pm$ 0.03         & 1.16 $\pm$ 0.08         \\
$R_{\star}$ ($R_{\rm \odot}$)       & 1.2  $\pm$ 0.1        & 1.02 $\pm$ 0.02         & 1.27 $^{+0.10}_{-0.05}$ & 0.90 $\pm$ 0.02         & 1.44 $\pm$ 0.05         \\ \noalign{\smallskip}
$\log g_{\star}$ (cgs)              & 4.26 $\pm$ 0.08       & 4.44 $\pm$ 0.02         & 4.28 $^{+0.03}_{-0.07}$ & 4.53 $\pm$ 0.02         & 4.18 $\pm$ 0.02         \\ \noalign{\smallskip}
$\rho_{\star}$ ($\rho_{\rm \odot}$) & 0.54$\pm$ 0.18        & 0.99 $\pm$ 0.05         & 0.56 $^{+0.06}_{-0.12}$ & 1.38 $\pm$ 0.10         & 0.39 $\pm$ 0.03         \\ \noalign{\smallskip}
$M_{\rm P}$ ($M_{\rm Jup}$)         & 1.23 $\pm$ 0.08       & 0.60 $\pm$ 0.07         & 0.28 $\pm$ 0.04         & 1.0 $\pm$ 0.1           & 1.16 $\pm$ 0.09         \\ \noalign{\smallskip}
$R_{\rm P}$ ($R_{\rm Jup}$)         & 1.4 $\pm$ 0.2         & 1.24 $\pm$ 0.03         & 0.96 $^{+0.10}_{-0.05}$ & 0.93 $\pm$ 0.03         & 1.21 $\pm$ 0.05         \\ \noalign{\smallskip}
$\log g_{\rm P}$ (cgs)              & 3.2  $\pm$ 0.1        & 2.95 $\pm$ 0.05         & 2.83 $\pm$ 0.09         & 3.42 $\pm$ 0.05         & 3.26 $\pm$ 0.04         \\
$\rho_{\rm P}$ ($\rho_{\rm J}$)     & 0.5 $\pm$ 0.2         & 0.32 $\pm$ 0.04         & 0.31  $\pm$ 0.08        & 1.2 $\pm$ 0.2           & 0.66 $\pm$ 0.07         \\
$a$ (AU)                            & 0.0363 $\pm$ 0.0007   & 0.0449 $\pm$ 0.0007     & 0.0449 $\pm$ 0.0008     & 0.0628 $\pm$ 0.0007     & 0.0345 $\pm$ 0.0007     \\
$e$                                 & $ < 0.058$            & $< 0.017 $              & $<0.18$                 & $<0.096$                & $<0.17$                 \\
$T_{\rm eq}$ (K)                    & $1600 \pm 80$         & $1400 \pm 30$           & $1480 \pm 60$           & $1100 \pm 25$           & $1790 \pm 40$          \\
$\sigma_{\rm jit}    $ (km s$^{-1}$)& 0.0225,0.0205         & 0.0115, 0               & 0.062, 0.014            & 0.0165                  &                        0\\
\hline 
\end{tabular}
\\ 
\end{table*}

\begin{table*}
\caption{Kinematics and distances. Proper motions are from
\citet{2013yCat.1324....0S}. \label{uvwtable}}
\begin{tabular}{lrrrrrr}
\hline
Star & 
\multicolumn{1}{l}{$\mu_{\alpha}$ [mas/yr]} &
\multicolumn{1}{l}{$\mu_{\delta}$ [mas/yr]} &
\multicolumn{1}{l}{d [pc]} &
\multicolumn{1}{l}{U [km\,s$^{-1}$]} &
\multicolumn{1}{l}{V [km\,s$^{-1}$]} &
\multicolumn{1}{l}{W [km\,s$^{-1}$]} \\
\hline
\hline
WASP-119&$22.1\pm2.1$&$ 16.0\pm1.8$&333$\pm$29&$ 28\pm4$&$-10\pm3$&$12\pm3$\\
WASP-124&$-2.0\pm1.3$&$-14.4\pm1.0$&433$\pm$11&$-16\pm2$&$-16\pm2$&$12\pm2$\\
WASP-126&$62.0\pm1.1$&$ 52.9\pm0.9$&234$\pm$15&$ 64\pm5$&$-47\pm3$&$14\pm2$\\
WASP-129&$11.8\pm1.3$&$  0.7\pm1.8$&246$\pm$~7&$-27\pm2$&$ -0\pm1$&$18\pm2$\\
WASP-133&$ 9.3\pm1.5$&$ -7.8\pm1.0$&547$\pm$21&$ 24\pm3$&$ -8\pm3$&$ 2\pm3$\\
\hline
\end{tabular}
\end{table*}

\section{Discussion}
 WASP-119 is a typical hot Jupiter system in terms of the mass (1.2~$M_{\rm
Jup}$), radius (1.3~$R_{\rm Jup}$) and orbital period (2.5\,d) of the planet.
The host star has a similar mass to the Sun but appears to be much older based
on its effective temperature and density (Fig.~\ref{trho_plot}).  

 WASP-124 is also a typical hot Jupiter system (0.6~$M_{\rm Jup}$, 1.2~$R_{\rm
Jup}$, 3.4\,d) but with an apparently much younger host star, although not so
young that it has an enhanced lithium abundance. The projected rotation
velocity of WASP-124\,A is one of the largest in our sample ($v\,\sin I
\approx 3$ km\,s$ ^{-1}$). Rapid rotation is expected for young solar-type
stars, but it is unclear whether the age estimates for planet host stars based
on their rotation rate (gyrochronological ages) are reliable
\citep{2015A&A...577A..90M}. 

 WASP-126\,A is the brightest star presented here and also hosts the
lowest-mass planet in our sample (0.3~$M_{\rm Jup}$). The radius of this
planet is quite large (0.95~$R_{\rm Jup}$) so it also has the lowest surface
gravity of these newly-discovered planets. This combination of low surface
gravity (i.e., large atmospheric scale height) and a bright host star make this
a good target for transmission spectroscopy. 
 
 The high density of WASP-129\,A can be explained using stellar models with
enhanced initial helium abudance. The helium-enhanced models we have used here
($\Delta \rm Y = +0.02$) provide an adequate fit to the observed properties of
this star,  so we expect that models with an even higher helium abundance
($\Delta \rm Y \approx +0.04$) would provide a much better fit.
\cite{2015A&A...577A..90M} found similarly high densities for two other hot
Jupiter host stars (HAT-P-11\,A and WASP-84\,A). These high densities inferred
from the transit light curves cannot be explained by contamination of the
light curve by a third star in the system since this would lead to a shallower
eclipse depth and a lower inferred value for the stellar density
\citep{2003ApJ...585.1038S}. The distribution of the initial helium abundance
for solar-type stars is very uncertain since there is no direct way to
measure the helium abundance of cool stars, so it is unclear how this
parameter impacts on our understanding of the age and mass distributions for
planet host stars.  The surface gravity of WASP-129\,A derived from our
analysis of the light curves and spectroscopic orbit (4.53 $\pm$ 0.02) is
clearly inconsistent with the value derived from our analysis of its spectrum
(4.1 $\pm$ 0.1). Analysis of the spectrum of WASP-129\,A at high resolution
and higher signal-to-noise may help to find the reason for this discrepancy.
WASP-129\,b has a slightly longer orbital period than most hot Jupiters
(5.7\,d) and, for the stellar mass that we have adopted, also has a smaller
radius (0.9~$R_{\rm Jup}$)  than most hot Jupiters with similar masses
(1.0~$M_{\rm Jup}$). The high density estimate for WASP-129\,b implied by
these parameters should be treated with some caution until the mass of the
host star can be determined with greater confidence.  The surface gravity of
this planet is also high compared to other hot Jupiters and this quantity is
independent of the assumed stellar mass.

 WASP-133\,A appears to be an old, metal-rich G-dwarf when compared to stellar
evolution models (Fig.~\ref{trho_plot}). WASP-133\,b is a hot Jupiter (1.2~$M_{\rm Jup}$, 1.2~$R_{\rm Jup}$) with the
shortest orbital period of the systems presented here (2.2\,d).
 The surface lithium abundance of WASP-133\,A 
(A(Li)=2.7) is significantly higher than other stars of similar age
and effective temperature \citep[A(Li) $<2.5$,][]{2005A&A...442..615S}. This
is a counter-example of the general trend for planet host stars to be depleted
in lithium compared to similar non-planet host stars
\citep{2010PASP..122.1465M,  2015MNRAS.446.1020G, 2015A&A...584A.105D}.
Whether this anomaly is connected to the slow rotation of this star or some
other factor is difficult to ascertain from the consideration of a single
system. Understanding the relationships between these different properties of
planets and their host star can only be done by studying a large sample of
planetary systems, such as the one being compiled using WASP-South of which
these new discoveries are the latest contribution.

\section*{Acknowledgements}
WASP-South is hosted by the South African Astronomical Observatory and we are
grateful for their ongoing support and assistance. Funding for WASP comes from
consortium universities and from the UK's Science and Technology Facilities
Council. TRAPPIST is funded by the Belgian Fund for Scientific  Research (Fond
National de la Recherche Scientifique, FNRS) under the  grant FRFC
2.5.594.09.F, with the participation of the Swiss National  Science Fundation
(SNF).  M. Gillon and E. Jehin are FNRS Research  Associates. The authors
thank the anonymous referee for helpful comments which have improved the
quality of this paper.

\bibliographystyle{aa} 
\bibliography{short}

\end{document}